\documentclass[preprint,aps]{revtex4}
\usepackage{graphicx}
\usepackage{dcolumn}
\usepackage{bm}

\newcommand{\dket}[1]{ \left| #1 \right\rangle }
\newcommand{\dbra}[1]{ \left\langle #1 \right| }

\newcommand{\ba}[1]{\begin{array}{#1} }
\newcommand{\ea}{\end{array}}
\newcommand{\rt}{ \frac{1}{\sqrt{2}} }

\begin{document}
\title{Comparison of Star and Ring Topologies for Entanglement
Distribution}
\author{A. Hutton and S. Bose}
\address{Centre for Quantum Computation, Clarendon Laboratory,
    University of Oxford,
    Parks Road,
    Oxford OX1 3PU, England}

\begin{abstract}
We investigate the differences between distributing entanglement
using star and ring type network topologies.  Assuming
symmetrically distributed users, we asses the relative merits of
the two network topologies as a function of the number of users
when the amount of resources and the type of the quantum channel
are kept fixed. For limited resources, we find that the topology
better suited for entanglement distribution could differ from
that which is more suitable for classical communications.
\end{abstract}

 \maketitle

\section{Introduction}

In quantum information processing, entanglement \cite{revs} is a
particularly useful resource and has many applications such as
secret key distribution \cite{Ek}, teleportation \cite{ben} and
dense coding \cite{wies}. Recently, these quantum communication
protocols have been implemented \cite{impl}. It is imaginable that
in the future a large number of distant users would want to
engage in communicating with each other through quantum
protocols. To enable this to happen, such distant users will need
to share particles in maximally entangled states, irrespective of
noise in the entanglement distribution channels. Various schemes
have been put forward which could directly or indirectly help in
such distribution of entanglement
\cite{qdistr1,purf1,purf2,purf21,bhm,zuk,multswp,seriesswp,qreapt},
and have been experimentally demonstrated \cite{distrexpts}.
Given a certain physical distribution of users intending to
communicate quantum mechanically, one can connect them with
quantum channels to construct networks for the distribution of
entanglement. Networks for such entanglement distribution can
have different architectures depending on which users are linked
directly by quantum channels and which users are indirectly
linked through intermediate nodes.

\begin{figure}
\begin{center}
\includegraphics[width=4in]{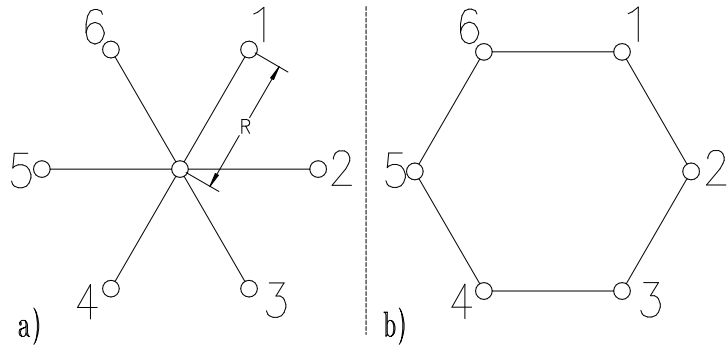}
\caption{This figure illustrates the two types of network
topologies considered, in this case for 6 parties. Each party is
located at a corner of  regular hexagon. The small circles denote
a party and connecting lines denote connecting quantum channels
through which entanglement is distributed.  R denotes the
distance of parties from the center.} \label{factors}
\end{center}
\end{figure}

 In classical networks, two of the major network topologies are the
star network and the ring network.  In a ring network, one
continuous ring joins all parties who wish to communicate (as
shown in Fig.\ref{factors}(a)), whereas in a star network all
parties are connected to a central hub where information is
exchanged (as shown in Fig.\ref{factors}(b)). If all the nodes of
such classical networks are assumed to be free from attacks and
failures, then wire length becomes the variable of interest for
comparison of the two network types - i.e., one network is said
to be better than the other when it requires less wire to
construct.  This is because noise in the connecting channels is
unimportant for classical communications. No matter how noisy the
connecting channels are, classical information can be amplified
arbitrarily and sent faultlessly through these channels. Based on
the wire-length criterion, for networks having the simple circular
layouts with symmetrically placed users, as shown in
Fig.\ref{factors}, the ring network is better than a star network
when the number of users is $N\geq 6$, while the reverse holds
true for $N<6$.

  For a quantum network, however, we use {\em entanglement}, rather
than wire-length, as a figure of merit in comparing networks.
This is due to the fact that when distributing entanglement via
quantum channels, unavoidable noise always degrades perfect
entanglement in the transmission process and consequently it is
not as easy to reliably distribute entanglement. We therefore use
the criterion that the better network is the one that permits the
sharing of a greater amount of entanglement between pairs of
users on average.

 To begin, we tackle the problem for a general
channel, when the available resources are unlimited or so large
that asymptotic entanglement distillation protocols
\cite{purf1,purf2,purf21} can be used as a part of the
entanglement distribution method. Note that physically such a
situation is permitted {\em only when} each user can ''store"
qubits noiselessly for a long time. This gives them the chance to
manipulate a large number of qubits together, as is required for
asymptotic entanglement distillation. In this asymptotic case, we
find the criterion for better entanglement distribution becomes
equivalent to the wire-length criterion of classical networks.
Next we examine two cases of extremely limited resources (only one
initial entangled pair available to one user, and exactly one
initial entangled pair available to each user) with a specific
quantum channel to illustrate the fact that the network for better
entanglement distribution can differ sharply from that for better
classical communications. After that, we give a heuristic
explanation of this striking difference between the cases of
limited and infinite resources as seen in our specific examples.

\section{A General Formulation of the Problem}

We consider a very simple type of network to facilitate study of
the problem.  Suppose there are $N$ parties, who are distributed
in a circle at a constant distance R from the centre, and wish to
share entanglement.  They connect themselves using quantum
channels using either a star or a ring layout (see
Fig.\ref{factors}).  As the number $N$ increases, the distance
between a party and its neighbors decreases.  We define a
'wirelength' as the shortest connection available on either
network - for the ring network this is the channel between
neighbors and for the star network it is the channel between one
party and the hub.

When any two parties wish to share entanglement, they must use
some distribution method to share entanglement between themselves
using the most efficient route available to them for the network
they are connected by.  For a given channel, this method will
involve sharing entangled pairs either directly between the two
interested parties or between intermediate parties and then
joining them using {\em entanglement swapping}
\cite{zuk,multswp,seriesswp,qreapt}. There may be some form of
distillation involved to concentrate the intermediate or final
entanglement in the distribution. In general, one will have to
adopt a specific entanglement distillation protocol. It could be
asymptotic or non-asymptotic, depending on the availability of
resources. It may be optimal or non-optimal depending on whether
the knowledge of the optimal distillation protocol exists for the
classes states generated during distribution through the channels
provided, and whether technology exists for its implementation. A
specific type of quantum channel and associated entanglement
distillation protocol (not necessarily optimal), together with
entanglement swapping to link up adjacent nodes, will comprise a
specific distribution method. The basic approach of this paper
will be to first choose a specific distribution method and then
compare its efficiency on the star and the ring lay-outs.  The
variables which describe the quantum channel available between
two parties are:

\begin{itemize}
\item $n$ - the number of wirelengths between them
\item $d$ - the length of the wirelengths between them
\end{itemize}

To clarify this, consider Fig. \ref{factors} which shows 6 parties
of whom two may wish to communicate.  If parties 1 and 4 wish to
share entanglement, then using the star network (Fig. 1(a)) they
must use $n=2$ wirelengths of length $d=R$ each.  If they are
connected using the ring network (Fig. 1(b)) then they must use
$n=3$ wirelengths of length $d=2R\sin(\frac{\pi}{6})$ each.

Let us define a function $E(d,n)$ which gives the entanglement
distributed between two parties separated by $n$ wirelengths each
of length $d$.  To be able to compare the two network layouts we
calculate an $E_{av}$ which is the distributed entanglement
averaged over all possible pairs of parties who wish to
communicate:

\begin{equation}
\label{generaleqn} E_{av} = \frac{1}{N-1} \sum
\limits_{n=1}^{N-1}E(d,n)
\end{equation}
For a star network $d$ is always $R$ and $n$ is always $2$ so we
have:

\begin{equation} E_{av}^{star} = E(R,2) \end{equation}
For a ring network, $d$ is the distance between two neighboring
parties, and is $2R\sin(\frac{\pi}{N})$ so we have:

\begin{equation} E_{av}^{ring} = \frac{1}{N-1} \sum \limits_{n=1}^{N-
1}E\left(2R\sin\left(\frac{\pi}{N}\right),n\right) \end{equation}
Bearing in mind that in a ring network entanglement can be
distributed either way round the network means this formula can be
refined to always use the shortest distance:

\begin{equation}
E_{av}^{ring} = \frac{1}{N-1} 2 \sum \limits_{n=1}^{U}
E\left(2R\sin\left(\frac{\pi}{N}\right),n\right) + \mu
E\left(2R\sin\left(\frac{\pi}{N}\right),N/2\right)
\end{equation}
where $U = (N-1)/2$ if $N$ is odd or $N/2 - 1$ if N is even and
$\mu$ is 0 if $N$ is odd, or 1 if $N$ is even.

To find at what point one layout becomes better than the other,
we are interested in finding, for a particular distribution
method, the $n$ where:

\begin{equation}
E_{av}^{ring} = E_{av}^{star}
\end{equation}
We are interested in comparing how this value of $N$ compares with
the classical case, where, for parameters we specify, the ring
network became better than the star network as the number of
parties is increased.  For a classical network, as noted before,
this occurs at $N=6$.

Unfortunately there is no analytic formula available for the
function $E(d,n)$ for an arbitrary quantum channel.  This is
because $E(d,n)$ represents the amount of entanglement that can
be distilled from a state after decoherence during transmission
through a noisy channel. No general formula is known yet for the
distillable entanglement of a given state. Despite this fact, it
is possible (as we will show) to provide a general statement
about the case when an unlimited number of pairs are available
across each wirelength. However, when only a limited amount of
resources are available per user, we have to rely on the explicit
form of the distillable entanglement. In case of limited
resources, we will therefore calculate $E(d,n)$ for specific
circumstances (specific channel types and specific distribution
methods) and use in Eq.(\ref{generaleqn}). The cases we consider
are:

\begin{itemize}
\item Distributing an unlimited number of pairs along each wirelength and linking each wirelength up using entanglement
swapping. We consider a {\em general} noisy channel for this case.
\item Distributing one pair traveling from source to destination
along one or more wirelengths. We use a specific type of quantum
channel and a specific distribution method.
\item Distributing one pair between each party, and then using
entanglement swapping to link them up. We use a specific type of
channel for this case as well.
\end{itemize}

We then try to investigate and explain the trends observed in the
above specific cases.


\section{The case of an unlimited number of pairs for a general channel}

First we consider the case where the parties in the network share
a very large number $T$ of maximally entangled pairs.  An equal
number $T/N$ is given to each party.  In the case of a ring
network each party then sends one half of the pair to their
neighbour on the left whereas in a star network each party would
send one half to a central hub.  The consequence is that in
either case, $T/N$ pairs are shared across each wirelength. Since
$T$ was very large, $T/N$ is very large (assuming $N$ stays
small, of course) and so an asymptotic number of maximally
entangled pairs can be distilled across each wirelength i.e.
$(T/N)E_D$ where $E_D$ is the distillable entanglement of a pair
that has decohered on travel through the wirelength.  In this
asymptotic case we assume that the maximally entangled states can
be collected together in each wirelength and then {\em matched}
(i.e., aligned end to end) with maximally entangled states in
adjacent wirelengths. The ability to do this would depend on
being able to discriminate and store the distilled maximally
entangled states. Connecting up adjacent maximally entangled pairs
in succession by entanglement swapping then produces $(T/N)E_D$
maximally entangled pairs shared between any two users. In this
asymptotic case, the same entanglement arises independent of the
number of wirelengths separating two users. In
Eq.(\ref{generaleqn}) then $E$ becomes a function of only the
length $d$ of a single wirelength. Therefore because $E$ depends
only on $d$, the situation becomes equivalent to the classical
case and the crossover at which the ring network becomes better
than the star network is also at $N=6$. This is because for $N=6$
a radial wirelength has the same length as a circumferential
wirelength and therefore $E_D$ will be the same for states
travsersing the star or ring networks.  For $N<6$ the radial
wirelengths are shorter than circumferential ones and so $E_D$
will be greater for states passing through a star network. For
$N>6$ then circumferential wirelengths are shorter and so the
ring network gives a greater overall entanglement.

\section{One Pair Travelling All The Way For A Bit-Flip Channel}

In this scenario we suppose that only one entangled pair is
provided to any one of the users and he may have to communicate
with any of the other users through a bit-flip channel. This means
that the two parties must distribute the single pair between
themselves. In the case of a star network one party sends one
half of the pair to a central point and then on again to the
other party.  For a ring network the half of the pair would
travel around the ring through other parties until it reached the
destination party. This channel acts on states in the following
way:

\begin{equation}
\dket{\psi^+}\dbra{\psi^+} \rightarrow \Lambda= \lambda
\dket{\psi^+}\dbra{\psi^+} + (1 - \lambda)
\dket{\phi^+}\dbra{\phi^+}
\end{equation}

We relate $\lambda$ to the length of the channel $d$ by:

\begin{equation}
\lambda = \frac{1+e^{-d}}{2}
\end{equation}

For such states the maximum distillable entanglement is known
\cite{vp} to be $1-S(\Lambda)$ where $S(\Lambda)$ denotes the Von
Neumann entropy of the state $\Lambda$.

\begin{figure}[ht]
\begin{center}
\includegraphics[width=4in]{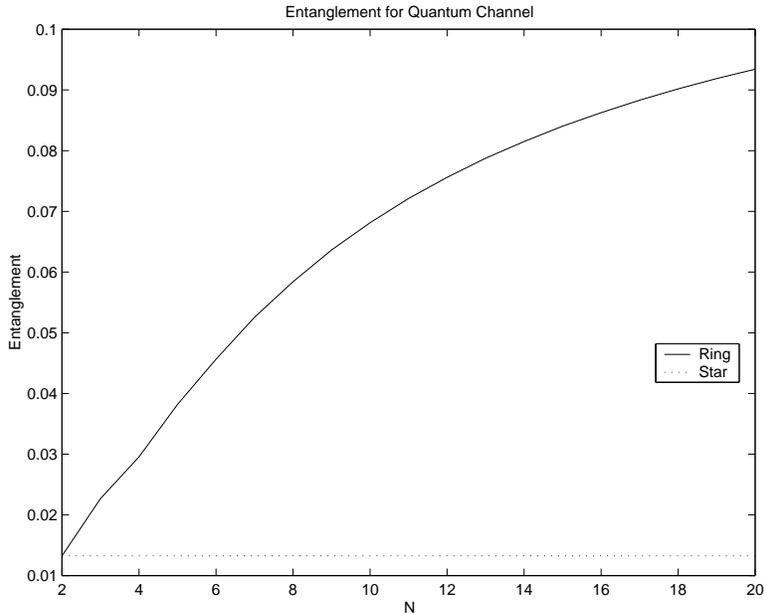}
\caption{This figure plots the average entanglement for a ring
network (solid line) and a star network (dotted line) against the
number of users in the network when only one entangled pair is
available to travel the entire distance between parties intending
to communicate. This graph is true for all values of the radius
$R$ and the channel type is bit-flip.} \label{oneonly}
\end{center}
\end{figure}

One particle was kept at the originator while the other was sent
to the other party, passing through one or more wirelength to
reach it. Fig.\ref{oneonly} shows $E_{av}^{ring}$ and
$E_{av}^{star}$ plotted against $N$.  We see that right from the
start ($N=2$) the ring layout is better. This is true for all
values of the radius $R$ of the network.  So we see a difference
from the classical case where it was only at $N=6$ that the ring
network became better than the star network.

\section{One Pair Between Each Party For A Bit-Flip Channel}

In this scenario we consider each party in the network having one
maximally entangled pair.  Each party then sends one particle of
the pair to their neighbour through a bitflip channel.
Entanglement swapping is then used to create a link between any
two parties who wish to share entanglement.  The procedure of
entanglement swapping produces a pair linking across the two
original pairs with a fidelity given by:

\begin{equation}
F_{n} = 1 + 2F_aF_b - F_a - F_b
\end{equation}

\begin{figure}[ht]
\begin{center}
\includegraphics[width=4in]{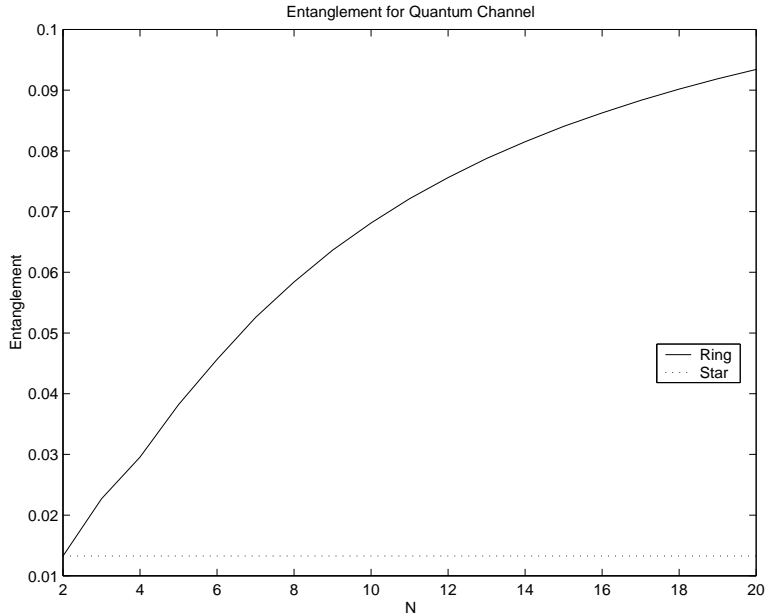}
\caption{This figure plots the average entanglement for a ring
network (solid line) and a star network (dotted line) against the
number of users in the network when only one entangled pair is
available in each wirelength. This graph is true for all values of
the radius $R$ and the channel type is bit-flip.} \label{oneeach}
\end{center}
\end{figure}

For a bit-flip channel, the order in which pairs are connected by
entanglement swapping does not matter.  Fig.\ref{oneeach} shows
$E_{av}^{ring}$ and $E_{av}^{star}$ plotted against $N$.  We see
again that straight away the ring layout is better. In fact, due
to the nature of the bit-flip channel, the resultant fidelities
for this situation and the previous one where one pair is shared
across the entire distance between the two communicating parties,
turn out to be identical.

Thus in the case dealt with in sections IV and V, it becomes
apparent that the results of comparing network topologies for
entanglement distribution can be \emph{very different} from the
classical case.

\section{A Heuristic Explanation Of The Results}

In this section, we attempt to  explain heuristically the
difference in the results for the unlimited number of pairs (where
the ring network becomes better than the star network only after
there are more than $6$ parties in the network), and the result
for one pair (where the ring network is always better than the
star network).

   We will first give a general description which encompasses both
the case for an unlimited number of pairs and that for a small
number of pairs between parties.  When a distillation procedure
operates on a finite ensemble \cite{finite}, we generally have
outcomes of various degrees of entanglement with various
probabilities.  In these outcomes, the entanglement is either
pumped up or pumped down from the original values. To try to
express a general pattern, we restrict ourselves to distribution
methods obeying the following assumptions:

1. Assume a general distillation protocol operating on
arbitrarily sized ensembles (finite or infinite) where there is a
probability $p$ that distillation in a wirelength is successful
and boosts the entanglement to $E_D + \delta_s$, $E_D$ being the
distillable entanglement of the state.  There is a probability
$1-p$ that it fails and the entanglement is reduced to $E_D -
\delta_f$.  In most cases there will be more than two possible
outcomes of a general distillation protocol, but for simplicity
we assume just two outcomes.

2. Assume $E_D + \delta_s$ is quite near to maximal so that the
entanglement swapping to connect adjacent wirelengths is near perfect.  When
adjacent pairs are connected using entanglement swapping we assume that the
resultant entanglement is equal to the lower value of the two pairs.
Therefore when connecting $n$ wirelengths, the entanglement will be the lowest
value of the $n$ wirelengths.

Using these rules, we can formulate the following expression for
the average entanglement obtained over $n$ wirelengths:

\begin{equation}
p^n (E_D + \delta_s) + (1-p^n)(E_D - \delta_f) = E_D +
p^n(\delta_s + \delta_f) - \delta_f.
 \label{eqn:ong}
\end{equation}
As we move from distributing a finite number of entangled pairs
to an infinite number the $\delta$'s will tend to zero and
$p\rightarrow 1$, meaning the average entanglement will
asymptotically tend to approach the distillable entanglement.  To
make it more clear, in the asymptotic case we have a unit
probability ($p \rightarrow 1$) conversion of a homogenous
ensemble to an inhomogenous ensemble of maximally entangled and
unentangled pairs with the fraction of maximally entangled pairs
being $E_D$.  This fraction $E_D$ of maximally entangled pairs in
each wirelength can now be connected with unit efficiency using
entanglement swapping. Asymptotic distillation essentially
conserves the distillable entanglement and tiny fluctuations
$\delta$ in the final entanglement tend to zero.  If we put $p
\rightarrow 1$ and $\delta \rightarrow 0$ in Eq.(\ref{eqn:ong}) we
are left with with an average of
 $E_D$ over different lengths and the
merit of the star and the ring networks simply depends on the
total wirelength.  In the other extreme (non asymptotic), the
fluctuation $\delta$ is large (say of the order $~E_D$) and we compare $\sum_{n}
p^n$ with $p^2$.  This leads to the behaviour of ring always
being better than the star.  Thus we have successfully interpolated between
the case of distributing a finite number of entangled pairs to an infinite
number.

   Note that in the above presentation, asymptotic distillation
   has been viewed in a rather different angle than usual. Usually
$E_D$ is interpreted as the probability of a successful
distillation and the entanglement produced as a result of a
successful distillation is maximal. We invert the interpretation
of this same process as a method succeeding with probability
$p=1$ and creating an amount of entanglement $E_D$ per initial
impure pair in the form of maximally entangled pairs. In this
latter (maximally entangled) form, the pairs can be connected in
series without any loss of entanglement through entanglement
swapping.

In order to illustrate the meaning of the above heuristic
approach with specific values of $p$ and $\delta$'s we consider an
example - a watched amplitude damped channel where we distribute
the state $\frac{1}{\sqrt{2}}\left( \dket{00} + \dket{11}
\right)$, with two pairs between connected users to obtain a case
somewhat intermediate between the previous examples.

\begin{equation}
\ba{lll}  |\Psi^{'}\rangle_{12E_1E_2}&=&\rt( \sqrt{ 1 + e^{-4d}}[
\frac{ \dket{00} + e^{-2d} \dket{11}}{ \sqrt{ 1 + e^{-4d}} }
]_{12} \dket{00}_{E_1E_2} \\ &+& e^{-d}\sqrt{1-e^{-2d}}
\dket{01}_{12}\dket{10}_{E_1E_2} \

\\
&+& e^{-d}\sqrt{1-e^{-2d}} \dket{10}_{12}\dket{01}_{E_1E_2} \\&+&
(1-e^{-2d}) \dket{00}_{12}\dket{11}_{E_1E_2}) \ea
\end{equation}
Then, if the environment is being monitored, there is a
probability of $\frac{1}{2} \left( 1+e^{-4d}\right)$ that the
state observed is (corresponds to the state $\dket{00}_{E_1E_2}$
of the environment)

\begin{equation}
|\Psi^{'}_c\rangle_{12}=\frac{1}{ \sqrt{1+e^{-4d}} } \left(
\dket{00} + e^{-2d} \dket{11} \right) \label{eq:c}
\end{equation}
where the subscript $c$ represents the fact that this state is a
result of conditional evolution. $|\Psi^{'}_c\rangle_{12}$ is not
maximally entangled and must now be purified.  One method of doing
this is to use the Procrustean method.  This has probability of
producing a maximally entangled stated (MES) of twice the modulus
squared of the lower coefficient. i.e.

\begin{equation}
\frac{2 e^{-4d} }{1 + e^{-4d}}
\end{equation}
We combine the probability of observation with that of
purification to give an expression for $p$ i.e. the successful
concentration to an MES.

\begin{equation}
p = e^{-4d}
\end{equation}
Before the entanglement was $E_D$, now it is 1.  So

\begin{equation}
\delta_s = 1 - E_D
\end{equation}
\begin{equation}
\delta_f = E_D
\end{equation}
So the expression for the average entanglement above becomes:

\begin{equation}
e^{-4nd}
\end{equation}
With this expression for one pair having been distributed, the
ring network is immediately better than the star network as
before.

\section{Conclusions}

 In this paper, we have compared entanglement distribution between several
users connected by star and ring network configurations. We have
shown that the cross over point at which the ring network becomes
better than the star network varies with the amount of resources.
When the amount of resources is limited and cannot be stored, so
that the number of entangled pairs available across a channel at a
time is finite, then the results for entanglement distribution
can differ sharply from that for classical communications. We
have given a heuristic explanation of this fact that it stems from
the comparison of different powers of probabilities arising in
the comparison of star and ring networks. However, in the
asymptotic case (which can physically arise when we can store
particles for long and can process a large number of shared
entangled pairs parallelly) the relative merits of the star and
the ring configurations are the same as classical.

 We have arrived at our conclusions by considering extreme examples.
 At one extreme is just one pair per wirelength and just one pair available
 to travel all the way from user to user. At the
 other extreme is a very large number of pairs shared in parallel
 between each adjacent user (or user and node) in which asymptotic manipulations of
 entanglement are used. As we gradually increase the number of pairs
 available to be processed parallelly between users
from $1$ to $\infty$, we would expect the cross over point in $N$
to increase from $1$ to $6$. This is because when more and more
pairs can be stored, we have a greater chance of selectively
connecting the higher entangled pairs in adjacent wirelengths.
When the number of pairs becomes really large, this selective
connection becomes really successful and gives just the
entanglement in a single wirelength as the effective criterion
for comparison. In the future, we intend to investigate explicit
examples when an intermediate number of pairs are shared in
parallel per wirelength to explore the transition from the
non-asymptotic to the asymptotic case.

\section{Acknowledgements}

AH thanks the UK EPSRC (Engineering and Physical Sciences
Research Council) for financial support.

\end{document}